\documentclass{article}
\usepackage{emulateapj,graphics}

\lefthead{BAKER ET AL.}
\righthead{THE ORS-PREDICTED VELOCITY FIELD}

\newcommand\svec{{\bf s}}
\newcommand\vvec{{\bf v}}
\newcommand\yre{{\cal Y}}
\newcommand\km{{\rm\ km}}
\newcommand\s{{\rm\ s}}
\newcommand\kms{{\rm\km\s^{-1}}}
\newcommand\Mpc{{\rm\ Mpc}}
\newcommand\iras{{\sl IRAS\/}}

\begin{document}

\title{The Velocity Field Predicted by the 
       Optical Redshift Survey}

\author{Jonathan E. Baker and Marc Davis}
\affil{Astronomy Department, University of California, Berkeley, CA 94720; \\
       jbaker@astro.berkeley.edu}

\author{Michael A. Strauss\altaffilmark{1}}
\affil{Princeton University Observatory, Princeton, NJ 08544}

\author{Ofer Lahav}
\affil{Institute of Astronomy, Cambridge University, Madingley Road, 
       Cambridge CB3 OHA, \\ United Kingdom}

\and

\author{Bas\'\i{}lio X. Santiago}
\affil{Departmento de Astronomia, Universidade Federal do Rio Grande do Sul,
       91501-970, Porto Alegre, RS, \\ Brasil}

\altaffiltext{1}{Alfred P. Sloan Foundation Fellow, and Cottrell Scholar of
		 Research Corporation}

\begin{abstract} 
We have used the Optical Redshift Survey (ORS; \cite{san95}) to construct the
gravity field due to fluctuations in the galaxy density field out to distances
of $8000\kms$.  At large scales where linear theory applies, the comparison of
this gravity field with the observed peculiar velocity field offers a powerful
cosmological probe, because the predicted flow field is proportional to the
parameter $\Omega^{0.6}/b$, where $\Omega$ is the matter density and $b$ is the
bias of the galaxy distribution.  The more densely sampled ORS gravity field,
to excellent approximation, matches that of the earlier \iras\ 1.2-Jy redshift
survey (\cite{fis95a}), provided $\beta$ is reduced by a factor
$b_{opt}/b_{IRAS} \approx 1.4.$  Apart from this scaling, the most significant
difference between the ORS and \iras\ fields is induced by differing estimates
of the over-density of the Virgo cluster.  Neither of these gravity fields is
consistent with the peculiar velocity field constructed from the full Mark III
(\cite{wil97a}) sample.  We find that a simple but plausible non-linear bias
algorithm for the galaxy distribution relative to the mass has a negligible
effect on the derived fields.  We conclude that the substitution of optical for
\iras\ catalogues cannot alone resolve the discrepancies between the \iras\
gravity field and the Mark III peculiar velocity field.
\end{abstract}

\keywords{cosmology --- dark matter --- galaxies: clustering ---
    large-scale structure of universe}


\section{Introduction}

With the advent of large, uniform redshift surveys of galaxies, increasingly
stringent tests of models for the large-scale structure of the universe have
become possible (see, e.g., \cite{dek94} and \cite{str95} for reviews).  Under
the assumption that large-scale flows of galaxies in the universe are a
response to the underlying distribution of matter, peculiar velocity
measurements are a critical probe of cosmology and large-scale structure.
Redshift surveys allow one to construct an estimate of the gravity field based
on the observed distribution of galaxies.  Using linear or quasi-linear
gravitational instability theory, one can predict the velocity field from this
gravity field, a prediction which is approximately proportional to the
parameter $\beta \equiv f(\Omega)/b$, where $b$ is the galaxy bias, $f$ is a
function which is well approximated by $\Omega^{0.6}$ (\cite{pee80}), and
$\Omega$ is the matter density.

Until recently, the Infrared Astronomical Satellite (\iras) 1.2-Jy flux-limited
redshift survey was the only available nearly full-sky catalogue.  The original
redshift survey of \iras-selected galaxies was limited to a sample of 2658
galaxies with 60-micron flux greater 1.936 Jy (\cite{str92a}), but this was
later extended to a sample of 5321 galaxies brighter than 1.2 Jy
(\cite{fis95a}).  A sample of 13,000 galaxies complete to 0.6 Jy is nearing
completion (\cite{can98}) and is eagerly awaited.  The principal advantage of
these \iras-selected samples is their insensitivity to extinction within the
Milky Way, allowing extremely large sky coverage, $87.6$\% (11.01 sr) for the
1.2-Jy sample, with a single, linear instrument.  Nearly full-sky coverage is
essential for estimation of the peculiar gravity field.

However, it is well known that the distribution of \iras\ galaxies is biased
with respect to the overall distribution of galaxies (\cite{str92b}), primarily
because \iras\ under-counts the dust-free early-type galaxies which congregate
in cluster centers.  The \iras\ surveys are also more dilute than is desirable,
typically including only 1/3 of the known spiral galaxies in volumes where the
selection function is high.  Optically-selected surveys can eliminate this
bias, and deeper redshift surveys will overcome the shot noise problem
associated with the sparse sampling of the \iras\ survey.  The recently
completed Optical Redshift Survey (ORS; \cite{san95}) is currently the best
available catalogue for our peculiar velocity analysis.  The ORS is a
concatenation of three optically-selected samples covering most of the sky with
$|b| > 20^\circ$, and it is the best current approximation to a full-sky
optically-selected catalogue.  Until a redshift catalogue is constructed from
the 2MASS survey (\cite{ste95}), the ORS is likely to remain the closest
competitor to the \iras\ full-sky catalogue.  Previous work (\cite{fre94})
suggests that the \iras\ and optical gravity fields are consistent, but with
the superior data now available, it is clearly of interest to determine the
nature of any deviations between the gravity fields of the optical and \iras\
samples, as these could substantially affect comparisons to observed peculiar
velocity fields.  Direct density field comparisons (\cite{san92};
\cite{str92b}) between the ORS and other redshift surveys will be carried out
in detail by \cite{str98}.

Comparisons of the gravity field derived from the \iras\ survey with observed
peculiar velocity samples have been carried out using a number of complementary
methods (see \cite{str95} for a review), with more recent studies performed by
\cite{dav96}, \cite{wil97b}, \cite{rie97}, \cite{sig98}, \cite{dac98}, and
\cite{wil98}.  These studies show that the \iras-predicted flow field is at
least qualitatively very similar to the measured velocity field, lending strong
support to the gravitational instability model for the growth of large-scale
structure.  Although the qualitative alignment of the \iras-predicted and Mark
III peculiar velocity fields is remarkably good, detailed quantitative
comparisons are, at present, less satisfactory.  DNW96 find large coherent
residuals between the \iras\ and Mark III fields, primarily in the dipole
component at large scales, which preclude a conclusive determination of
$\beta$.  \cite{wil97b} find better alignment of the \iras\ gravity and Mark
III velocity fields using the rather different VELMOD technique, but this
analysis is limited to redshifts within $3000\kms$; \cite{wil98} find good
agreement on larger scales, but conclude that this requires a re-calibration of
the Tully-Fisher relations for the various subsamples going into the Mark III
catalogue inconsistent with the published calibration (\cite{wil97a}).  On
the other hand, \cite{dac98} find a better match between the large-scale \iras\
gravity field and the SFI I-band sample of spiral field galaxy peculiar
velocity measurements (\cite{gio97}).  The SFI sample uses the same
\cite{mat92} data for the Southern sky as in Mark III, but the transformation
of the data to a common system differs from the Mark III treatment.  A recent
summary of the current confusion is given by \cite{dav98}.  Might some of the
discrepancy disappear if we were to use an optical redshift survey to construct
an estimate of the gravity field?

This paper addresses the question of the construction of a three-dimensional
gravity field based on the ORS and \iras\ redshift catalogues.  We construct
our estimate of the gravity field using the redshift-space procedure described
by \cite{nus94}.  Section 2 describes the galaxy catalogues and our procedure
for merging them.  The method for obtaining the gravity and velocity fields is
briefly outlined in \S 3, with details of the spherical harmonic formalism in
the Appendix; more details have been presented elsewhere.  Section 4 describes
the differences between the ORS field and the field derived from \iras\ alone,
including a brief discussion of alternatives to the simple linear biasing
scheme, and \S 5 discusses the main conclusions.


\section{The Galaxy Catalogues}

The Optical Redshift Survey (ORS) is a combination of three individual
declination-limited catalogues: the Uppsala General Catalogue (UGC) at northern
declinations $(\delta\geq-2.5^\circ)$, the European Southern Observatory (ESO)
catalogue in the south $(\delta<-17.5^\circ)$, and the Extension to the
Southern Observatory Catalogue (ESGC) in the remaining strip just south of the
Celestial Equator.  The sample selection and galaxy distribution have been
described in detail (\cite{san95}), and the selection functions for the
galaxies in each sample have been carefully calibrated (\cite{san96}).  We use
the flux-limited versions of the UGC and ESO catalogues; for the ESGC
catalogue, only a diameter-limited sample was available.  All velocities have
been converted to the Local Group (LG) frame using the transformation of
\cite{yah77}.

Hudson (1993a, 1993b) constructed a full-sky map of the local density field
from optically-selected galaxies, using the UGC and ESO catalogues limited to a
redshift of $8000\kms$.  Complete redshift information did not exist for these
catalogues at the time of his analysis, and his sky coverage was limited to
67\%.  Furthermore, his procedure for merging the UGC and ESO into a single
catalogue may have led to spurious asymmetries between the North and South.
The complete ORS redshift sample allows us to overcome many of these
limitations.

Our analysis relies on a spherical harmonic expansion of the density field and
therefore requires data of uniform quality over the entire sky.  Unfortunately,
there are several limitations which prevent the ORS from being a uniform,
all-sky survey.  Because the ORS galaxies are optically selected, the catalogue
is affected much more adversely by extinction than is the \iras\ survey.  The
ORS does not cover the strip at Galactic latitudes $|b|<20^\circ$, while the
\iras\ survey excludes only the strip with $|b|<5^\circ$, two small unobserved
strips at higher latitudes, and small confusion-limited regions (about 12.4\%
of the sky in total).  In addition, we only use ORS data in regions where the
\cite{bur84} blue extinction is low: $A_B < 0.7$ magnitudes.  Finally, there
are 64 plates, each $5^\circ$ on a side, which are missing from the ESO-LV
survey (Fig.~5a of \cite{san95}).  The accepted ORS data cover about 62\% of
the sky, and the completeness of the catalogues within these regions is
excellent.

To produce an all-sky catalogue, we use the \iras\ survey to fill in the areas
in which ORS data are absent or the extinction is high.  The \iras\ catalogue
(\cite{fis95a}) contains 5321 galaxies covering 87.6\% of the sky.  We use a
version of this catalogue which \cite{yah91} have carefully augmented with fake
galaxies in the Zone of Avoidance (and other excluded regions) without biasing
the statistics of the galaxy distribution.  This all-sky \iras\ catalogue
contains 6010 galaxies.  The \iras\ data are only used in regions where there
are no accepted ORS data, so each of the four surveys cover disjoint regions on
the sky.

The ORS survey has the advantage that it is much more densely sampled than the
\iras\ survey within $cz\sim 8000\kms$.  This allows the local gravity and
velocity fields to be constructed with much smaller sampling errors.  However,
the ORS selection function drops exponentially with distance, while the \iras\
galaxy distribution has a long power-law tail extending to higher redshifts.
Beyond $8000\kms$, the ORS data become unacceptably sparse.  We therefore use
only \iras\ data over the whole sky for redshifts $cz>8000\kms$, extending the
analysis to $cz=18,000\kms$.

To generate a density field in redshift space, we first merge the galaxy
catalogues into a single, all-sky, uniform catalogue.  Each catalogue $j$ has a
selection function $\phi_j$ such that for homogeneously distributed galaxies,
the expected galaxy density at a point $\svec=(s,\theta,\varphi)$ is $\bar
n_j\phi_j(\svec)$, where $\bar n_j$ is the true average density of galaxies in
the catalogue volume.  Since we are given the selection functions $\phi_j$
(\cite{san96}), we calculate the density in (some part of) the volume of
catalogue $j$ using
\begin{equation}
\bar n_j = \frac{1}{V_j} \sum_{i\in j} \frac{1}{\phi_j(\svec_i)},
\end{equation}
where the sum is over galaxies $i$ in the volume of interest.  The sky coverage
$\Omega_j$ of each of the surveys is known (\cite{san95}), so 
\begin{equation}
V_j = \frac{1}{3} \Omega_j s_{max,j}^3
\end{equation}
is known.  Galaxies with $s_i>s_{max,j}$ are not included in the analysis.  We
calculate the density within $s_{max} = 8000 \kms$ for the three ORS
catalogues and the \iras\ catalogue, although, as mentioned above, the
\iras\ analysis is carried out to $18,000 \kms$. 
The densities we compute are for objects sufficiently luminous to be
included in a volume-limited sample out to radius $cz=500\kms$.  

The fractional density fluctuation is given by the ratio of the observed galaxy
counts to the number expected given the selection function.  This can be
calculated by assigning each galaxy $i$ a weight $\left[ \bar n_j
\phi_j(\svec_i) \right]^{-1}$.  In our merged catalogue, we multiply these
weights by factors which normalize the density in each catalogue to the density
of \iras\ galaxies \emph{within the accepted volume of the optical subsample}.
This allows each ORS subsample to have a density which, due to inhomogeneities,
is not necessarily the same as the overall density, but it matches the
different ORS subsamples to the uniformly-selected \iras\ catalogue.  Thus we
calculate $\bar n_{IRAS,j}$, the density of \iras\ galaxies within the volume
of catalogue $j$, relative to $\bar n_{IRAS}$, the overall density of \iras\
galaxies.  The weight for any galaxy is then given by
\begin{equation}
W_i = \frac{1}{\bar n_j\phi_j(\svec_i)} \frac{\bar n_{IRAS,j}}{\bar
n_{IRAS}}.
\label{eq:Wi} 
\end{equation}
This weighting ensures that the density of the merged catalogue remains
approximately constant across catalogue boundaries.  Some parameters of the
catalogues are listed in Table \ref{cats} (cf.\ \cite{san96}).

\begin{table*}
\begin{center}
\begin{tabular}{ccccr@{.}lr@{.}lr@{.}l}
Catalogue & \multicolumn{1}{c}{Sky} & \multicolumn{1}{c}{$N_{gal}$} & 
\multicolumn{1}{c}{$N_{gal}$} & \multicolumn{2}{c}{$\bar n_j$} & 
\multicolumn{2}{c}{$\bar n_{IRAS,j}$} & 
\multicolumn{2}{c}{$\bar n_{IRAS,j}/\bar n_{IRAS}$} \\
 & \multicolumn{1}{c}{coverage} & \multicolumn{1}{c}{(total)}
& \multicolumn{1}{c}{(used)} & \multicolumn{4}{c}{$(h^3\Mpc^{-3})$} 
& \multicolumn{2}{c}{ } \\ 
\tableline
\iras\ & 38\% & 6010 & 3020 & 0&0488 & 0&0488 & 1&047 \\
UGC  & 34\% & 3246 & 2992 & 0&0799 & 0&0473 & 1&015 \\
ESO  & 19\% & 2412 & 2187 & 0&109  & 0&0361 & 0&775 \\
ESGC &  9\% & 1351 & 1203 & 0&195  & 0&0570 & 1&223 \\
\end{tabular}
\end{center}
\caption{Numbers of galaxies, densities, and catalogue weights for the
    catalogues used in the ORS density field analysis.  The \iras\ values
    include only the portion of the survey outside the accepted ORS volumes;
    the \iras\ sky coverage is 100\% for $cz>8000\kms$.  The full-sky \iras\
    density for $cz<8000\kms$ (optimal for minimum-variance estimates of the
    density) is $n_{IRAS}=0.0466\ h^3\Mpc^{-3}$.
\label{cats}}
\end{table*}


\section{Calculation of the Velocity Field}

\cite{nus94} show that in linear theory, the peculiar velocity field in
redshift space is irrotational and can therefore be derived from a potential:
$\vvec(\svec) = -\nabla\Phi(\svec)$.  If the angular dependences of the
potential and galaxy density field $\delta^g\equiv\delta\rho^g/\bar\rho^g$
(both measured in redshift space) are expanded in terms of spherical harmonics
as described in Appendix A, the potential is related to the density field by
\begin{eqnarray}
\lefteqn{ \frac{1}{s^2}\frac{d}{ds}\left(s^2\frac{d\Phi_{lm}}{ds}\right)
    - \frac{1}{1+\beta} \frac{l(l+1)\Phi_{lm}}{s^2} = } \nonumber \\
& & \frac{\beta}{1+\beta} \left(\delta^g_{lm} 
    - \frac{1}{s}\frac{d\ln{\phi}}{d\ln{s}}\frac{d\Phi_{lm}}{ds}\right).
    \label{peq}
\end{eqnarray}
This is a modified Poisson equation, where $\delta^g_{lm}$ is the spherical
harmonic coefficient of the redshift-space galaxy density field, $s$ is the
redshift, and $\beta$ was defined above.  The last term on the right hand side
corrects for the ``rocket effect'', which arises because we weight each galaxy
by the selection function $\phi$ evaluated at its redshift rather than its
(unknown) distance.

We now briefly summarize our computationally efficient method for solving the
above equation.  We first compute a smoothed density field on a spherical grid
in redshift space.  We construct a $32\times 32$ angular grid equally spaced in
Galactic longitude and latitude, with 52 bins in redshift out to $18,000\kms$,
providing more than adequate resolution given the smoothing of the fields.  The
separation of the redshift bins increases in proportion to the mean \iras\
inter-particle spacing, $(\bar n\phi)^{-1/3}$, in order to reflect the
decreased sensitivity of the surveys at high redshift.  The Gaussian-smoothed
galaxy density field at grid point $n$ is given by
\begin{equation}
1 + \delta^g(\svec_n) = \frac{1}{(2\pi)^{3/2}\sigma_n^3} 
    \sum_{i} W_i\ e^{-(\svec_n-\svec_i)^2/2\sigma_n^2}, 
\end{equation}
where the sum is over all galaxies in the merged catalogue.  The Gaussian
smoothing width for the cell, $\sigma_n$, is equal to the mean \iras\
inter-particle spacing at that redshift (or $100\kms$ when the inter-particle
spacing is smaller than this).  The increased smoothing at larger redshifts is
essential to prevent divergence of the increasing shot noise; our procedure is
qualitatively close to the optimal Wiener filtering procedure (\cite{lah94};
\cite{fis95b}), and the result is a signal-to-noise ratio in the density field
that is roughly constant over radial bins.  A correction factor must also be
applied because part of the Gaussian-weighted volume centered on a cell will
fall outside the survey cutoff radius.

The spherical harmonic coefficients of the density field are calculated on
radial (redshift) shells.  Once these coefficients $\delta^g_{lm}$ have been
computed for the merged catalogue, the modified Poisson equation for the
potential is solved.  The velocity field can then easily be constructed from
$\vvec = -\nabla\Phi$.

Our analysis is done purely in redshift space, assuming linear theory and a
one-to-one mapping between distance and redshift along any given line of sight.
These assumptions become invalid in clusters, where highly nonlinear,
virialized motions cause a radial distortions into ``fingers of God.'' We
follow \cite{yah91} in collapsing the galaxies in the six nearest clusters of
their Table 2 to a single redshift.

A limitation of the density field constructed by merging disparate catalogues
is that we can only use one selection function for the $d\ln\phi/ds$ term on
the right hand side of the modified Poisson equation (see Appendix B for
details).  The maps below were constructed using the \iras\ selection function,
which, however, is not a very good approximation to the ORS selection functions
for $cz>8000\kms$.  To gauge the importance of the differing catalogue
selection functions in the rocket effect correction, we have compared the ORS
peculiar flow field using both $\phi_{UGC}$ and $\phi_{IRAS}$.  (Differences
among the three ORS selection functions are negligible in comparison to the
difference between ORS and \iras.)  Although the rocket effect correction is
substantial over the redshifts of interest, the $\phi_{UGC}$ and $\phi_{IRAS}$
flow fields are identical (to within $\sim 30\kms$, with negligible systematic
differences over the sky) for $cz < 8000\kms$, where the selection functions
behave similarly.  At greater distances, however, the fields diverge
dramatically, particularly in the dipole.  Because we use only \iras\ data for
$cz>8000\kms$, we are fortunately able to avoid this complication.


\section{Comparison of the Fields}

A detailed statistical comparison of the ORS and \iras\ gravity fields is
difficult due to the limited sky coverage of the ORS.  Because we require an
all-sky catalogue for the spherical harmonic analysis, we have filled in the
Galactic plane with \iras\ galaxies; the two fields are therefore not
independent over the whole sky.  In addition, the ORS shares some galaxies in
common with the \iras\ survey.  Nevertheless, the qualitative comparison of the
two fields can tell us whether or not improved redshift surveys might provide a
better match to the Mark III data than DNW96 found for the \iras\ survey.
Detailed Wiener-filtered reconstructions of the \iras\ density and gravity
field in \emph{real} space have been carried out (\cite{web97}; see also
\cite{str95}); our analysis, however, is done entirely in redshift space.

\subsection{ORS and IRAS Density Fields}

The density field contours on several radial shells are illustrated in Figure
\ref{smooth}, where it is apparent that the ORS and \iras\ fields are
qualitatively very similar (compare with the similar figures in \cite{san95}).
The \iras\ survey's under-counting of cluster centers is most evident on the
near ($cz=1000\kms$) radial slice, which contains Virgo ($\ell=284^\circ$,
$b=75^\circ$), Ursa Major ($\ell=145^\circ$, $b=66^\circ$), and Fornax
($\ell=237^\circ$, $b=-54^\circ$).  With the Gaussian smoothing for this shell
($\sigma=345\kms$), the ORS over-density for Virgo is $\delta^g\approx 10$,
compared with $\delta^g\approx 6.7$ in the \iras\ field (recall that the
smoothing increases with radius, so density contrasts appear smaller on more
distant shells).  Note that if the \iras\ galaxies differed from the ORS
galaxies only by an extra bias factor of $2/3$, which would match the peak
Virgo over-densities for the two fields, then we would expect that a comparison
of observed peculiar velocities with the ORS gravity field would yield a value
of $\beta$ \emph{lower} than the corresponding number for the \iras\ gravity
field by the same factor.

On the next redshift slice ($cz=3000\kms$), the Hydra-Centaurus complex is
clearly visible around ($\ell=300^\circ$, $b=20^\circ$).  The maximum
over-densities are approximately 25\% smaller for \iras\ than for ORS.
However, the smoothing on this shell is somewhat larger ($\sigma=550\kms$), and
the clusters are quite near the ORS $b=20^\circ$ cutoff.  The linear structure
of the Supergalactic Plane is visible near $\ell=135^\circ$, and there are
pronounced voids ($\delta^g\approx -0.9$) towards $\ell=265^\circ$,
$b=-50^\circ$ and $\ell=5^\circ$, $b=10^\circ$.

At $cz=5000\kms$, the Perseus-Pisces region around ($\ell=140^\circ$,
$b=-25^\circ$) is clearly much better sampled by the ORS survey than by the
\iras\ survey, but the qualitative features of the two fields are identical.
The smoothing at this redshift is $815\kms$.  The $b<-20^\circ$ parts of the
Pavo-Indus-Telescopium region are also visible near $\ell=330^\circ$. The
density contrasts on more distant shells are much less pronounced, and beyond
$8000\kms$ the two fields are identical by construction.

\subsection{Predicted Flow Fields}

Figure \ref{vfield} shows the \iras- and ORS-predicted radial peculiar velocity
fields (Local Group frame) on redshift shells for $\beta=0.6$ (note these are
not the same as the redshift shells in Fig.~\ref{smooth}).  Again, the
overall agreement of the fields is quite good.  The nearby shell ($cz=500\kms$)
is dominated by Virgo-centric infall, with galaxies towards Virgo and galaxies
in the opposite direction of the sky flowing away from us.  The ORS-predicted
flow towards Virgo is approximately twice as large as the \iras-predicted flow
for the same value of $\beta$.  There are also substantial ($v\ga 300\kms$)
predicted flows towards us from the Supergalactic poles, but the differences in
these regions between the ORS and \iras\ predictions are more modest than for
the flow toward Virgo.

At $cz=2000\kms$, the dipole component of the flow begins to dominate as a
result of the motion of the Local Group, with infall on the back side of Virgo
and outward flow in the opposite direction.  The flow of galaxies away from us
as it falls into Hydra-Centaurus is modest but visible at this redshift.  The
dipole pattern grows with redshift, as evident in Figure \ref{vfield}c, and
except for an overall scaling and slight change of orientation, the differences
between the ORS and \iras\ fields diminish.

The monopole, dipole, and quadrupole terms as a function of radius for the ORS-
and \iras-predicted flow fields are shown in Figures 
\ref{monopole}--\ref{quadrupole} for the same value $\beta=0.6$, the high end
of the ``most likely'' range of DNW96.  The quantity plotted is $c_{lm}$, as
defined in Appendix A.  As emphasized by \cite{dav98} and \cite{nus94}, the
monopole and dipole terms at a given redshift are sensitive only to the
interior mass distribution, which should be well sampled.  Tidal fields
introduced by errors in the density field of the dilutely sampled exterior
region will only affect the quadrupole and higher multipoles.  Statistically
significant disagreement between the measured and predicted velocity field for
the dipole component is thus a sign that something is amiss, either a
systematic error in one or both of the catalogues, incorrect treatment of an
important component of non-linear flow, or non-uniform relative bias in the
galaxy distribution.  These figures compare the multipole moments of the \iras-
and ORS-predicted velocity fields, to see whether the DNW96 discrepancy
continues to hold for the ORS data.

The ORS-predicted dipole terms (Fig.~\ref{dipole}) are all significantly
larger than the \iras-predicted terms.  The differences in the Galactic $X$
($m=1$) and $Y$ ($m=-1$) components arise primarily behind Perseus-Pisces, but
the $Z$ ($m=0$) difference is mainly due to Virgo.  The quadrupole terms agree
quite well, except that the $m=-1$ magnitude is somewhat lower in \iras, and
the ORS data resolve the $m=0$ peak around Perseus-Pisces a bit more sharply.
We should point out that because each of the ORS subsamples are normalized to
the \iras\ density field (eq.~[\ref{eq:Wi}]), the \iras\ and ORS dipoles are
not as statistically independent as we would otherwise like them to be.
 
Figures \ref{monopole}--\ref{quadrupole} also show the results when the ORS
$\beta$ value is lowered from $0.6$ to $0.4$.  Note that equation \ref{peq}
is only linear in $\beta$ in the limit $\beta \rightarrow 0$.  For $\beta$ of
order unity, redshift distortions lead to more complex behavior; thus the
$\beta=0.6$ and $\beta=0.4$ ORS multipole terms are not simply proportional to
each other.  The overall amplitude of the ORS-predicted dipole for $\beta=0.4$
agrees extremely well with the \iras-predicted dipole for $\beta = 0.6$, 
but the increased weight of Virgo enhances the relative strength of the
$Z$-component.  Lowering the value of $\beta$ has a fairly modest effect on the
quadrupole terms, although the peak in the overall quadrupole amplitude at
$2500\kms$ is significantly reduced.  The $m=-1$ component is brought into
excellent agreement with the \iras\ prediction, and the $m=\pm 2$ amplitudes
are lowered somewhat below the \iras\ prediction.  In any case, the
incorporation of the ORS data has worked in the wrong direction to correct the
\iras--Mark III dipole residual of DNW96, because the $Y$ component of the Mark
III field is much too small relative to the \iras\ prediction, and the ORS data
only drive these components further out of agreement.

The dipole of the \iras-predicted velocity field does not seem to converge to
the CMB dipole ($\ell=268^\circ$, $b=27^\circ$ in the LG frame) particularly
well (\cite{str92c}; Webster et al.\ 1997); it consistently points about
$30^\circ$ away.  With their larger Virgo-centric infall, the ORS data only
exacerbate this situation.  At $cz=6000\kms$, the ORS dipole points to
$\ell=250^\circ$, $b=55^\circ$, about $15^\circ$ \emph{higher} in latitude than
the \iras\ dipole.  This is reflected in the larger magnitude of the $l=1, m=0$
multipole component of the field (Fig.~\ref{dipole}), which points in the
Galactic $Z$ direction.

\subsection{Predictions for Mark III Galaxies}

The predicted peculiar velocities for the Mark III galaxies in three redshift
slices are plotted in Figures \ref{vpec1000}--\ref{vpec5000}.  The \iras\ plots
differ from those in DNW96 only in the projection used and the orientation of
the longitude coordinate.  Note that the \iras\ predictions use $\beta = 0.6$
while ORS uses $\beta = 0.4$, consistent with the apparent relative bias of the
two fields of a factor of 1.5 (\cite{str92b}; \cite{her96}).  In general, the
difference between the ORS- and \iras-predicted fields is quite small.  In
fact, the magnitude of these residuals is generally lower than the residuals
seen by DNW96 in their tests on mock catalogues.  The strong velocity shear
seen in the Mark III sample across the Hydra-Centaurus complex is plainly
absent from either of the redshift surveys; indeed, this region shows only very
small differences between the fields.  The largest systematic differences are
seen for Perseus-Pisces; our peculiar motion away from this complex is smaller
in the ORS-predicted field.  The detailed statistical analysis of DNW96 shows
discrepancies between \iras\ and Mark III which would not be alleviated by our
ORS gravity field.

\subsection{Effects of Non-linear Bias}

In general, the bias of the galaxy distribution is the function which relates
fluctuations in the galaxy number density $\delta\rho^g$ to fluctuations in the
overall matter distribution $\delta\rho$.  The functional form depends on the
details of galaxy formation, which are poorly understood.  For simplicity, it
is usually taken to be a linear function independent of scale, so that the bias
parameter $b$ has a single numerical value.  Semi-analytic work (\cite{kau97};
see also \cite{wei95}) suggests that most physical bias mechanisms lead to a
nearly linear bias on large scales, but the value of the bias depends quite
strongly on the power spectrum, galaxy type, and galaxy luminosity in a given
sample.  This is of course problematic for flux-limited surveys, in which the
most distant galaxies are also the most luminous.  The ORS data in fact show
some evidence of a weakly scale-dependent bias (\cite{her96}). 

Provided that the bias is linear, our technique will recover the value of
$\beta\approx\Omega^{0.6}/b$.  Non-linear bias in the galaxy distribution will
introduce systematic errors in the analysis.  We investigate the impact of a
non-linear bias by postulating a gridded mass density fluctuation field $\delta
= \left[{\cal B}(\delta^g) - \bar{\cal B}\right]/b$, where ${\cal B}$ is some
non-linear function.  The mean value $\bar{\cal B}$ over the set of grid points
must be subtracted because the fluctuation field has zero mean by definition.

Most prescriptions for non-linear bias are functions of the local over-density
$\delta$.  With the variable smoothing of our computed density field, however,
it is difficult to implement such a non-linear bias in a spatially uniform
fashion.  But the mean value of the field is $\delta=0$ independent of
smoothing, and it is therefore possible to construct a non-linear bias using
$\delta=0$ as a pivot.  A simple non-linear form is the piecewise linear
function
\begin{equation} 
{\cal B}(\delta^g)=\left\{
\begin{array}{ll}
\delta^g & \delta^g>0 \\
(2/3) \delta^g & \delta^g \le 0
\end{array}
\right. \ ,
\end{equation}
as shown in Figure \ref{nonlin}.  This in effect makes galaxies more strongly
clustered than dark matter in over-dense regions, and less strongly clustered
in under-dense regions, relative to the case of linear bias.  In other words,
we are stealing dark matter from the clusters, filling the voids with more than
would be expected on the basis of the galaxy distribution.  The galaxy voids
are thus substantially deeper than the matter voids for reasonable values of
$b$ (Cen \& Ostriker 1992, 1993).  This is plausible (though admittedly ad hoc)
because galaxy formation is expected to be more efficient in dense regions, and
it is consistent with the results of simulations (\cite{whi87}; \cite{dek97};
\cite{lem98}).  We emphasize that our crude piecewise linear bias is only meant
to be illustrative; a more complex algorithm could not readily be implemented
because of the variable smoothing of our density field.

Even the fairly dramatic ``knee'' we have introduced in the bias at
$\delta^g=0$ has quite a modest effect on the predicted flow field.  The
overall amplitude of the flow is reduced, as expected because the dark matter
distribution is now less clustered.  However, the magnitude of this reduction
is only $\sim 10\%$ (see Fig.~\ref{nonlin}), and the best-fit value of
$\beta$ would therefore only increase slightly.  The direction of the dipole is
altered by only $\sim 2^\circ$, and other qualitative features of the flow
field are unchanged.  It seems quite unlikely that even a rather extreme
uniform bias prescription could plausibly eliminate the coherent residuals seen
between Mark III and our predicted velocity fields.


\section{Conclusions}

The gravity field derived from the ORS is quite similar to the field derived
from the \iras\ survey, provided the value of $\beta$ is multiplied by a factor
of $\approx 2/3$, roughly consistent with the relative strength of clustering
in the two surveys.  Some differences from this simple scaling are observed;
most significantly, the dipole of the ORS-predicted flow field is rotated
towards Virgo and is further than the \iras\ prediction from the CMB dipole.
This non-scaling behavior arises because the ratio of ORS to IRAS density
fluctuations is larger in the local Supercluster than in other regions.

DNW96 have found systematic discrepancies between the Mark III and
\iras-predicted flow fields, particularly a large coherent dipole residual,
which cannot be physical because the dipole field depends only on the interior
mass distribution.  Estimating the gravity field from the ORS survey has
provided a check on the influence of the known shortcomings of the \iras\
catalogue (namely, its sparse sampling and under-representation of cluster
cores).  Since the \iras\ and ORS gravity fields are so similar, we conclude
that the redshift-space catalogues are not themselves the source of the DNW96
discrepancies.

Compared to \iras, the ORS gravity field will generally push us towards lower
values of $\beta$ in order to fit a given sample of peculiar velocity
measurements.  The result of DNW96 for the \iras-Mark III comparison would
then suggest $\beta\la 0.45$, but this comparison with Mark III remains suspect
due to to the substantial quantitative discrepancies between the \iras\ and
Mark III fields.  \cite{dac98}, on the other hand, showed that the SFI sample
agrees quite well with the \iras\ gravity field for $\beta=0.6$; presumably
this sample would also be consistent with the ORS field for $\beta\approx 0.4$.
Another serious concern for the Mark III comparison is the fact that
complementary methods for attacking the velocity field problem have not
converged to a consistent, unambiguous result.  The POTENT analysis of
\cite{sig98} and \cite{dek93} favors larger values ($\beta\sim 1$) than the ITF
and VELMOD analyses.  These methods weight the data in very different ways, and
given that the fields show some level of inconsistency, it is not surprising
that different analyses of the same data yield different answers.

The sources of the DNW96 discrepancies between the \iras\ gravity field and the
Mark III peculiar velocity measurements remain uncertain.  These discrepancies
might be telling us that galaxies cluster in a way which is not closely related
to the clustering of the mass, or that the linear gravitational instability
model provides an inadequate description of flows on the scales of interest.
Alternatively, there may be unknown systematic biases in the Mark III data;
see also the discussion in \cite{wil98}.  We have investigated a simple but
plausible generalization of the usual linear bias model for galaxy clustering,
but this does not significantly affect features in the predicted velocity
field; even a rather sizeable change in the slope of the bias relation at
$\delta^g=0$ causes only a modest reduction in the amplitude of the field.
Improved full-sky peculiar velocity surveys (e.g., \cite{str97}) of larger
galaxy samples are eagerly awaited.  Such surveys are quite a difficult
undertaking, but they may go some way towards clarifying our understanding of
large-scale flows in the universe.

\acknowledgments

J.~E.~B. acknowledges support from an NSF graduate fellowship.  This work was
supported in part by NSF grant AST95-28340.  M.~A.~S. acknowledges the support
of the Alfred P. Sloan Foundation, Research Corporation, and NSF grant
AST96-16901.

\begin{appendix}
\section{Spherical Harmonic Formalism}

We solve for the linear peculiar velocity field from the density field expanded
in spherical harmonics on radial (redshift) shells.  The conventional spherical
harmonics are defined to be
\begin{equation}
Y_{l m}(\theta, \varphi) \equiv K_{l m} P_{l m}(\cos\theta)\ e^{im\varphi},
\end{equation}
where
\begin{equation} 
K_{l m} \equiv \left[ \frac{2l+1}{4\pi}
\frac{(l-m)!}{(l+m)!}\right]^{1/2} 
\end{equation}
and the $P_{l m}$'s are the associated Legendre polynomials.  Since the
density and velocity fields are real, it is convenient to work with the
real-valued spherical harmonic functions, which differ slightly from their
complex cousins.  Following \cite{bun96}, we define them as
\begin{equation}
\yre_{l m} \equiv (-)^m K_{l m} P_{l m}(\cos\theta) \times
   \left\{ \begin{array}{c@{\quad\quad}l}
   1 & m=0 \\
   \sqrt{2}\ \cos m\varphi & m > 0 \\
   \sqrt{2}\ \sin |m|\varphi & m < 0. \end{array}\right.
\end{equation}
The coefficients $a_{l m}$ of an arbitrary real field $A(\svec)$ are related
to the field by 
\begin{eqnarray}
a_{l m}(s) &=& \int d\Omega\ \yre_{l m} A(\svec) \\
A(\svec) &=& \sum_{l=0}^{\infty} \sum_{m=-l}^l a_{l m}(s)\yre_{l
m}.
\end{eqnarray}
It is convenient to normalize the coefficients of multipole order $l$ by the
value of $\yre_{l 0}$ at the North Galactic Pole; we therefore define
\begin{equation}
c_{l m} \equiv \left( \frac{2l+1}{4\pi} \right)^{1/2} a_{l m}.
\end{equation}
With this normalization, $c_{11}$ is the amplitude of the dipole
vector projected along the
Galactic $X$ axis ($\ell=b=0^\circ$), $c_{1,-1}$ is the amplitude along
the $Y$ axis ($\ell=90^\circ$, $b=0^\circ$), and $c_{10}$ is the amplitude
along the $Z$ axis ($b=90^\circ$).  The amplitude of the dipole is the
quadrature sum of the three coefficients.  The quadrupole terms of the
real-valued spherical harmonics are
\begin{eqnarray}
\yre_{20} &=& \frac{1}{4}\sqrt{\frac{5}{\pi}} (3\cos^2\theta - 1) \\
\yre_{21} &=& \frac{1}{4}\sqrt{\frac{15}{\pi}} \sin 2\theta \cos\varphi \\
\yre_{2,-1} &=& \frac{1}{4}\sqrt{\frac{15}{\pi}} \sin 2\theta \sin\varphi \\
\yre_{22} &=& \frac{1}{4}\sqrt{\frac{15}{\pi}} \sin^2\theta \cos 2\varphi \\
\yre_{2,-2} &=& \frac{1}{4}\sqrt{\frac{15}{\pi}} \sin^2\theta \sin 2\varphi,
\end{eqnarray}
where Galactic longitude is $\ell=\varphi$ and latitude is $b = 90^\circ -
\theta$.

\section{Rocket Effect Corrections}

A limitation of constructing a merged density field out of different catalogues
and then solving for the velocity field is that equation (\ref{peq}) only
allows for a single radial selection function in the rocket effect term
(\cite{kai87}).  This term corrects for the fact that, when constructing the
density field, we have evaluated the selection function $\phi$ at each galaxy's
redshift rather than its (unknown) distance.  This can be a substantial
correction to the redshift-space density field (\cite{fis95b}).

For the ORS survey, unfortunately, the selection function $\phi(\svec)$ is not
independent of angle on the sky, but depends on the catalogue and the
extinction along the line of sight.  The linear equation for the velocity
potential $\Phi$ of \cite{nus94} is
\begin{equation}
\frac{1}{f} \nabla^2\Phi + \frac{1}{s^2}\frac{\partial}{\partial s}
\left(s^2\frac{\partial\Phi}{\partial s}\right) = 
\frac{1}{b} \left( \delta^g - \frac{1}{s}\ \frac{\partial\ln\phi}{\partial\ln
s}\ \frac{\partial\Phi}{\partial s} \right), \label{rock}
\end{equation}
where $f\approx\Omega^{0.6}$, $b$ is the bias, and $\delta^g$ is the galaxy
over-density in redshift space.  A general selection function which is a
function of position on the sky $(\theta, \varphi)$ will introduce couplings
between different modes of the spherical harmonic expansion.  Expanding $\Phi$
on each redshift shell as
\begin{equation}
\Phi(\svec) = \sum_{l=0}^\infty \sum_{m=-l}^l \Phi_{lm}(s)
Y_{lm}(\theta,\varphi) 
\end{equation}
and similarly for $\delta^g$, multiplying through by $Y_{lm}^*$ and integrating
over solid angle, we find that the term subtracted from $\delta^g$ in equation
(\ref{peq}) becomes
\begin{equation}
- \frac{1}{s}\sum_{l^\prime=0}^\infty \sum_{m^\prime=-l^\prime}^{l^\prime}
u_{l^\prime m^\prime} \int_{(4\pi)} d(\cos\theta)\ d\varphi\
Y_{lm}^* Y_{l^\prime m^\prime} \frac{\partial\ln\phi}{\partial\ln s},
\end{equation}
where $u\equiv -\partial\Phi/\partial s$ is the radial peculiar velocity.  For
a selection function independent of azimuthal angle $\varphi$, this expression
reduces to
\begin{equation}
-\frac{2\pi}{s} \sum_{l^\prime\geq m} u_{l^\prime m}
K_{l m} K_{l^\prime m} \int_{-1}^1 d(\cos\theta)\ P_{l m}
P_{l^\prime m} \frac{\partial\ln\phi}{\partial\ln s}.
\end{equation}
This expression replaces the $-(u_{lm}/s)(d\ln\phi/d\ln s)$ term in equation
(\ref{peq}).  For our catalogues, the angular dependence of the selection
function is approximately just a step discontinuity at $|b|=20^\circ$, so by
symmetry, even (odd) values of $l^\prime$ make no contribution to odd (even)
values of $l$.  The utility of the spherical harmonic expansion breaks down
unless the $l^\prime=l$ term is dominant.  For the analysis at $cz<8000\kms$,
the corrections are negligible because the ORS and \iras\ selection functions
track each other quite well.  A deeper analysis, or one in which the selection
functions of the merged catalogues are not so similar, may need to take account
of these corrections. 

\end{appendix}

\newcommand{\rsplot}[4][0]{\begin{figure} \begin{center} \scalebox{#2}{
  \rotatebox{#1}{\includegraphics{#3}}} \end{center} \caption{#4}
  \end{figure}}

\clearpage

\rsplot[270]{0.85}{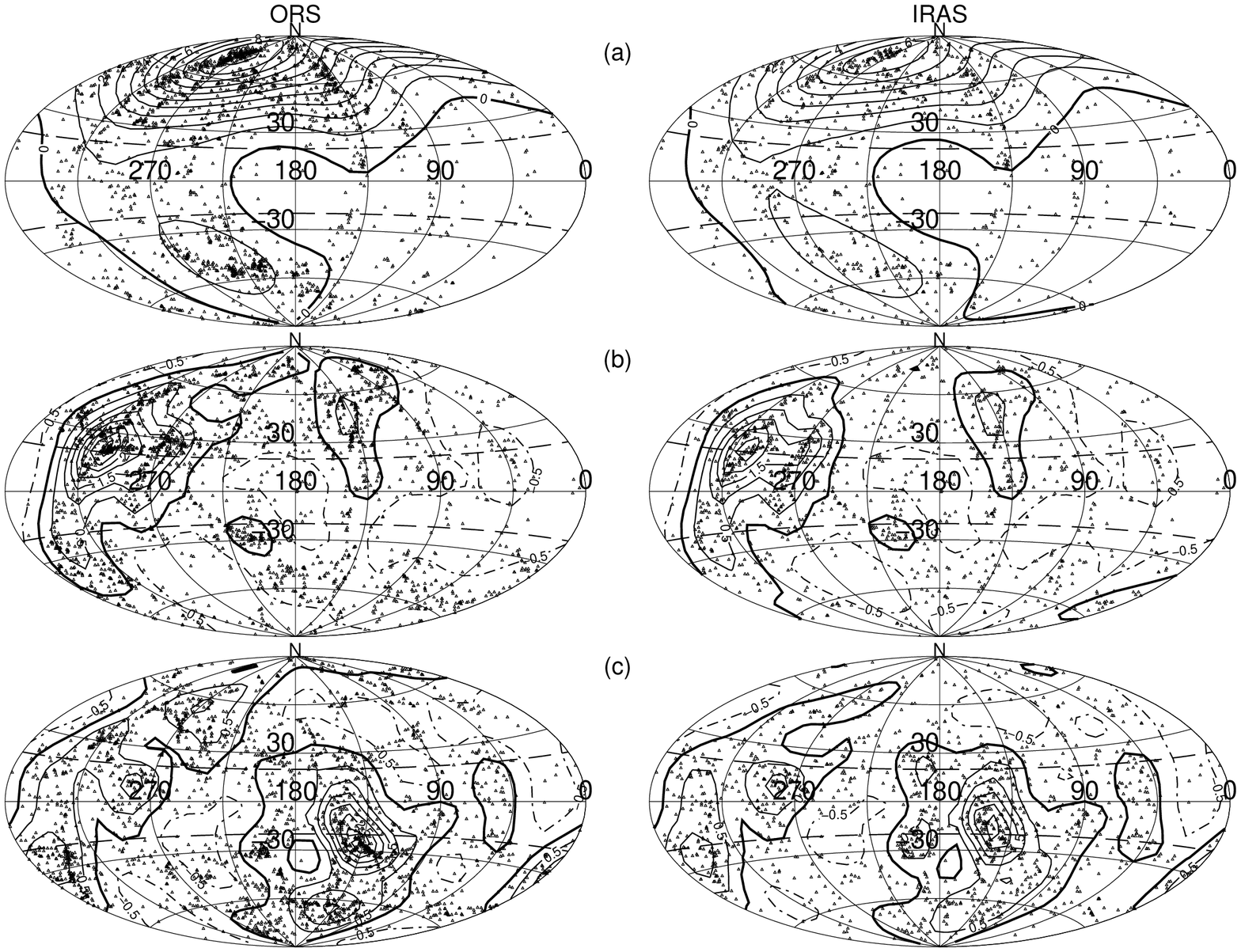}
{Smoothed density field contours on radial shells and catalogue galaxies in
redshift slices.  The Gaussian smoothing width subtends approximately
$10^\circ$ on each shell.  The long-dashed line shows the ORS $|b|=20^\circ$
cutoff.  All redshifts are measured in the Local Group frame.  (a) The nearby
shell $cz=1000\kms$ with galaxies in the range $0 \leq cz \leq 2000\kms$,
contour spacing 1.0, Gaussian smoothing width $\sigma=345\kms$.  (b) The
intermediate shell $cz=3000\kms$ with galaxies in the range $2000 \leq cz \leq
4000\kms$, contour spacing 0.5, Gaussian smoothing width $\sigma=550\kms$.  (c)
The distant shell $5000\kms$ with galaxies in the range $4000 \leq cz \leq
6000\kms$, contour spacing 0.5, Gaussian smoothing width $\sigma=815\kms$.
\label{smooth}}

\clearpage

\rsplot[270]{0.85}{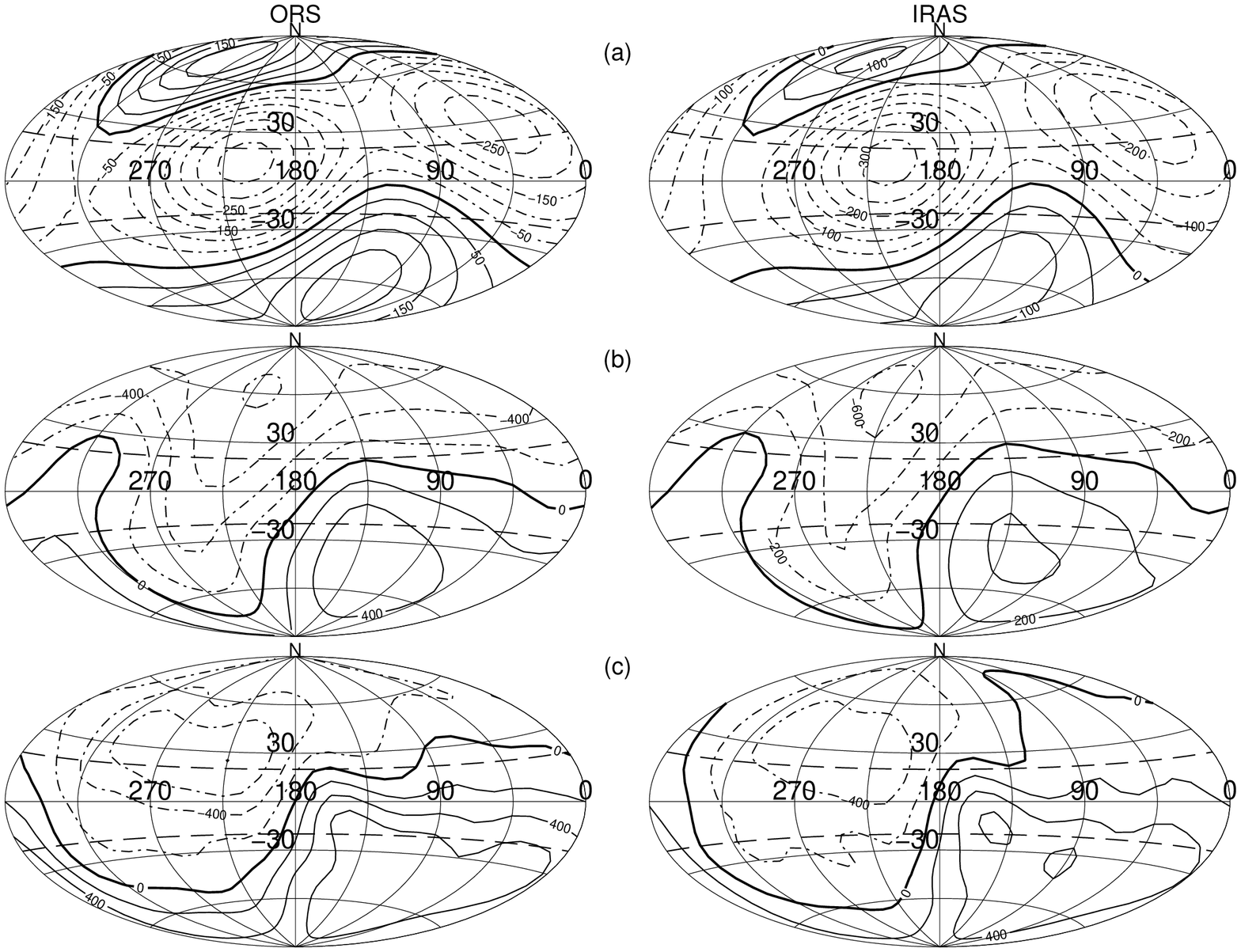}
{Radial slices of the ORS- and \iras-predicted peculiar velocity fields in the
Local Group frame for $\beta=0.6$.  Dot-dashed contours show negative
(approaching the Local Group) velocities.  (a) $cz=500\kms$, contour spacing
$50\kms$.  (b) $cz=2000\kms$, contour spacing $200\kms$.  (c) $cz=5000\kms$,
contour spacing $200\kms$.
\label{vfield}} 

\clearpage

\rsplot{0.9}{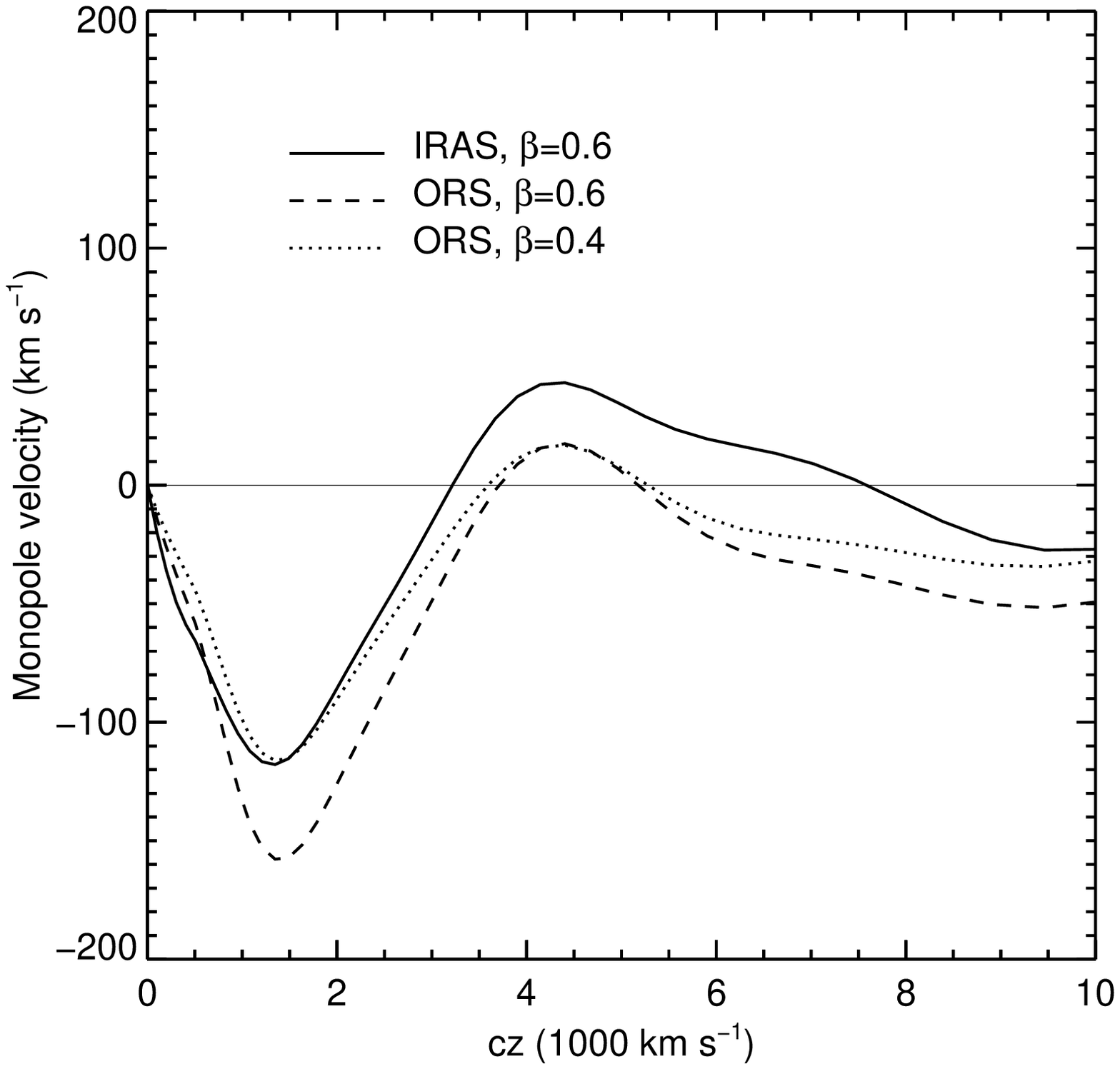}
{Monopole coefficient $c_{00}$ of the real-valued spherical harmonic expansion
of the ORS- and \iras-predicted peculiar velocity fields.  The solid line is
the \iras-predicted field for $\beta=0.6$, the dashed line is the ORS-predicted
field for the same $\beta$, and the dotted line is the ORS-predicted field when
$\beta$ is reduced to 0.4.
\label{monopole}}  

\clearpage

\rsplot{0.9}{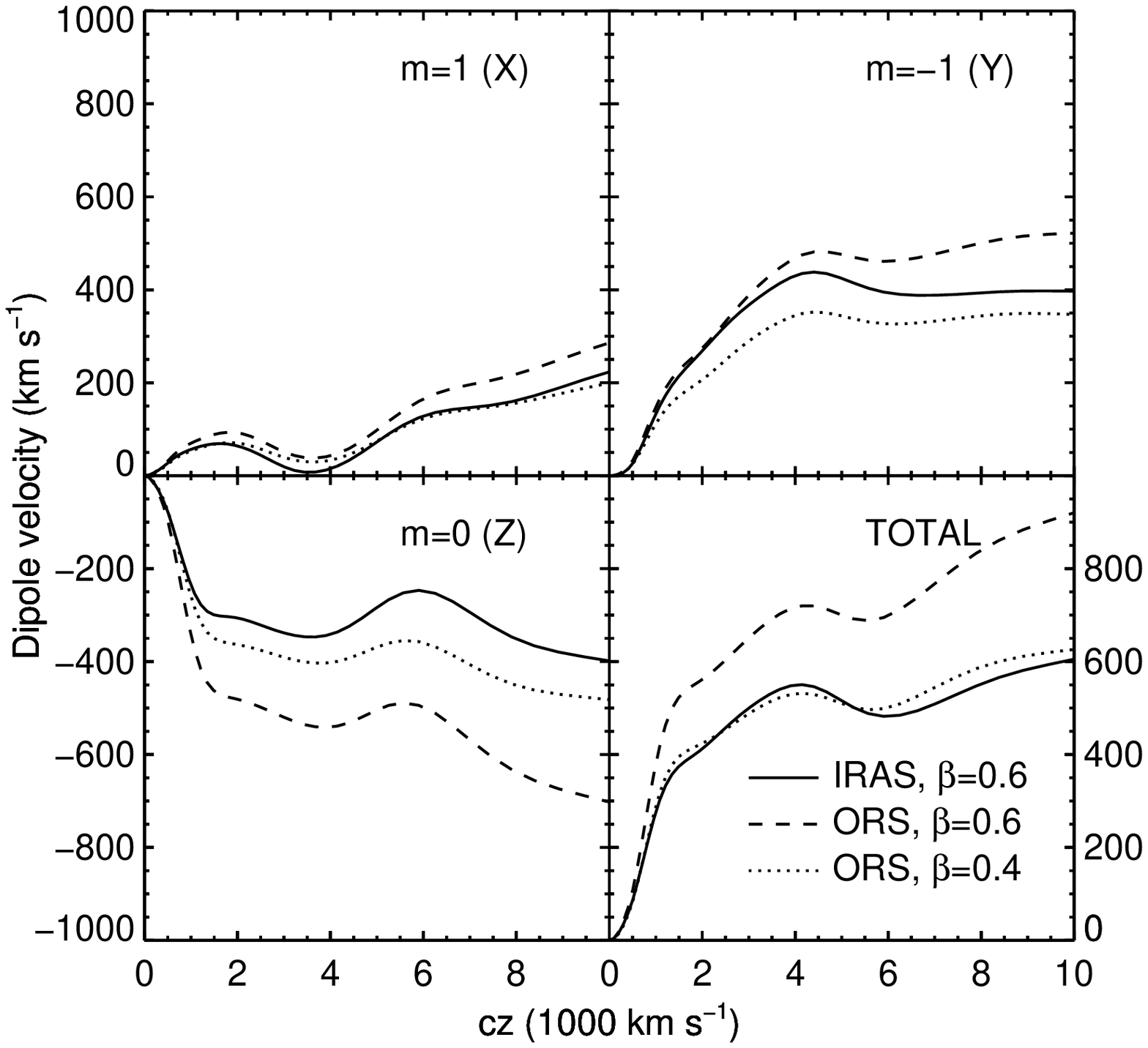}
{As in Figure \ref{monopole}, but for the dipole coefficients $c_{1m}$.  The
quadrature sum is shown in the lower right panel.
\label{dipole}}

\clearpage

\rsplot{0.9}{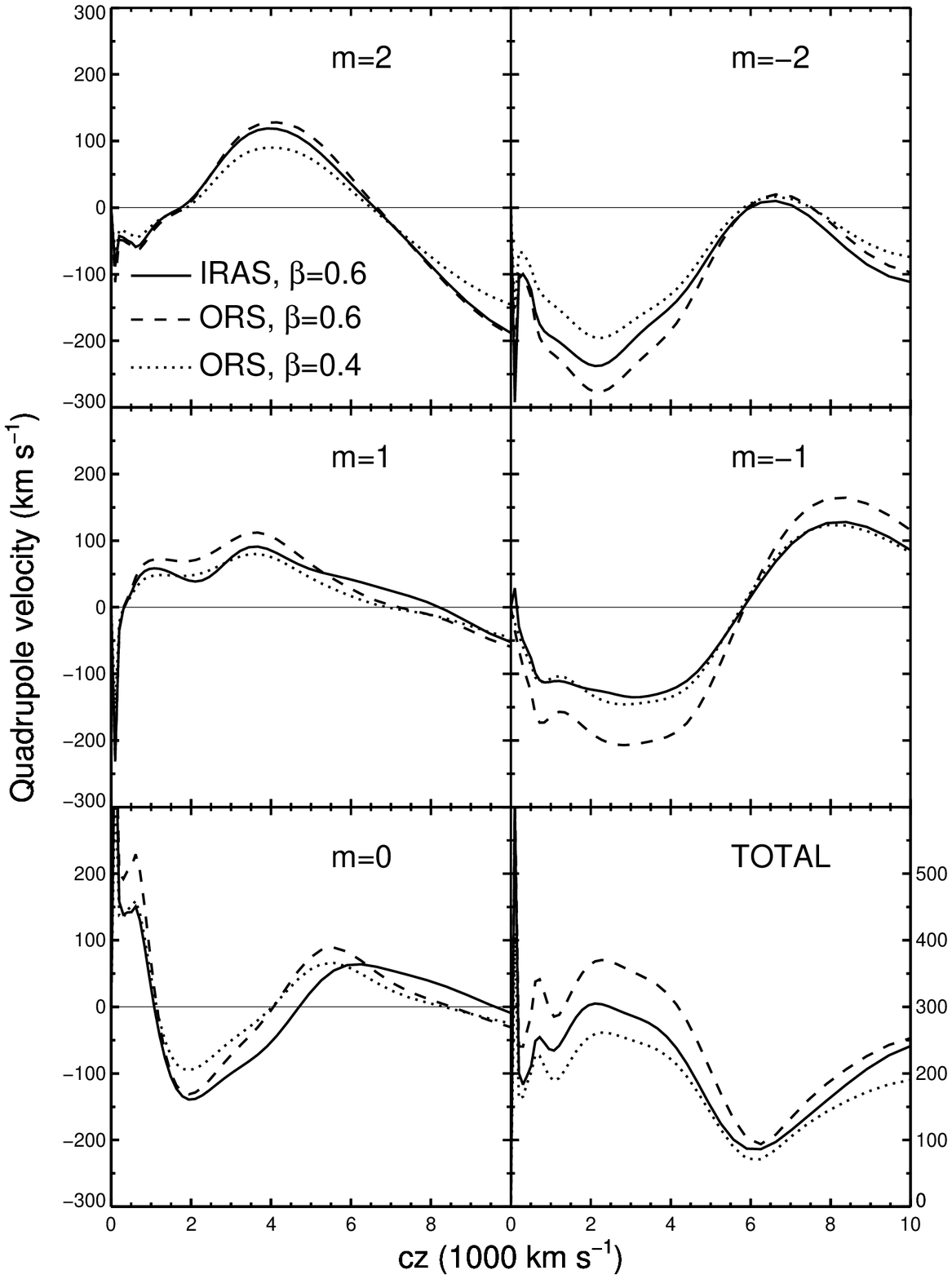}
{As in Figure \ref{monopole}, but for the quadrupole coefficients $c_{2m}$.
The quadrature sum is shown in the lower right panel.
\label{quadrupole}}

\clearpage

\rsplot{0.9}{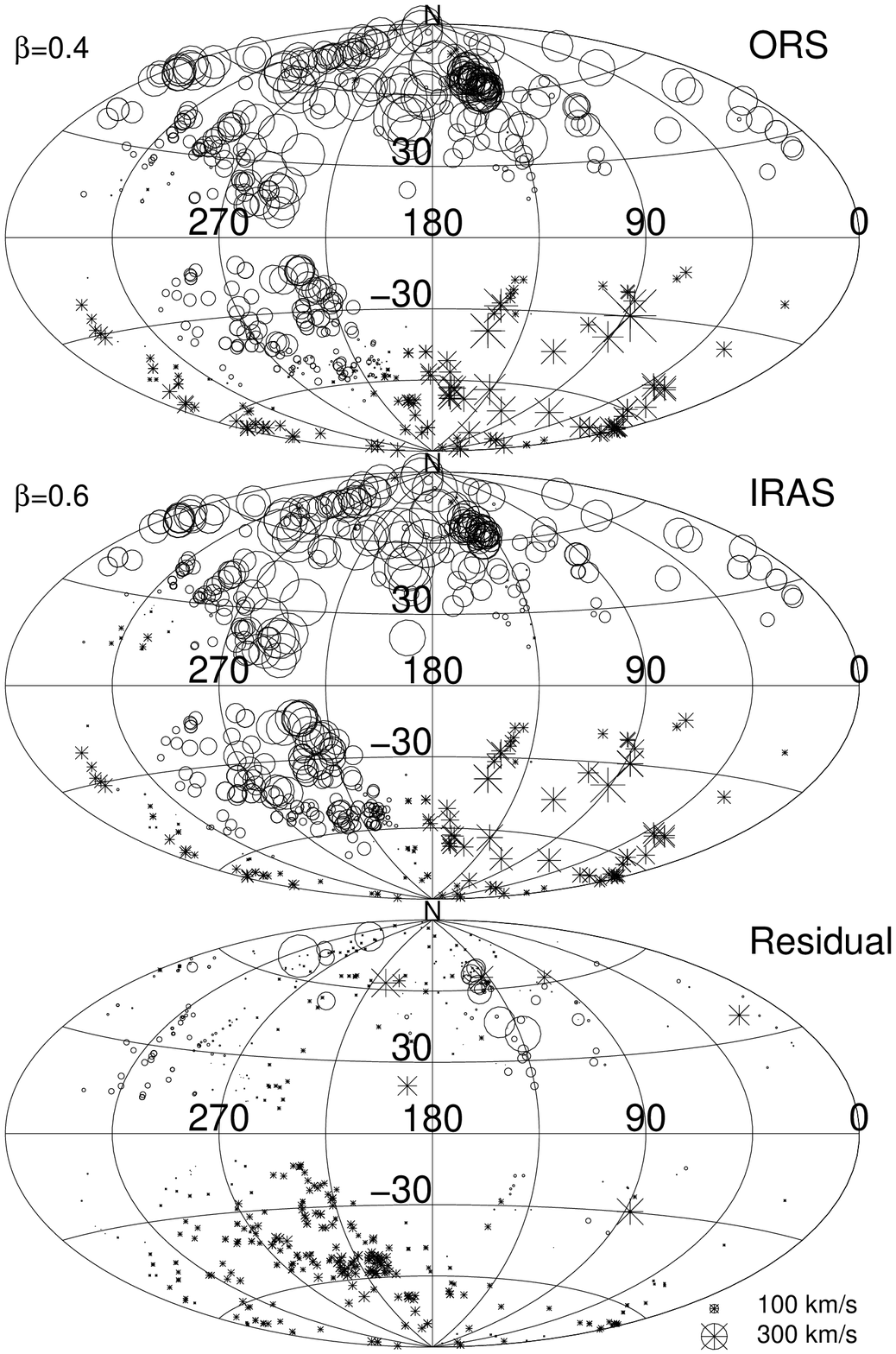}
{Predicted peculiar velocities for the Mark III galaxies based on ORS
($\beta=0.4$, top) and \iras\ ($\beta=0.6$, middle) for the near redshift slice
($cz<2000\kms$).  Circles indicate positive peculiar velocities (directed away
from the LG), and crosses indicate negative velocities.  The bottom plot shows
the difference between the ORS and \iras\ predictions.
\label{vpec1000}} 

\clearpage

\rsplot{0.9}{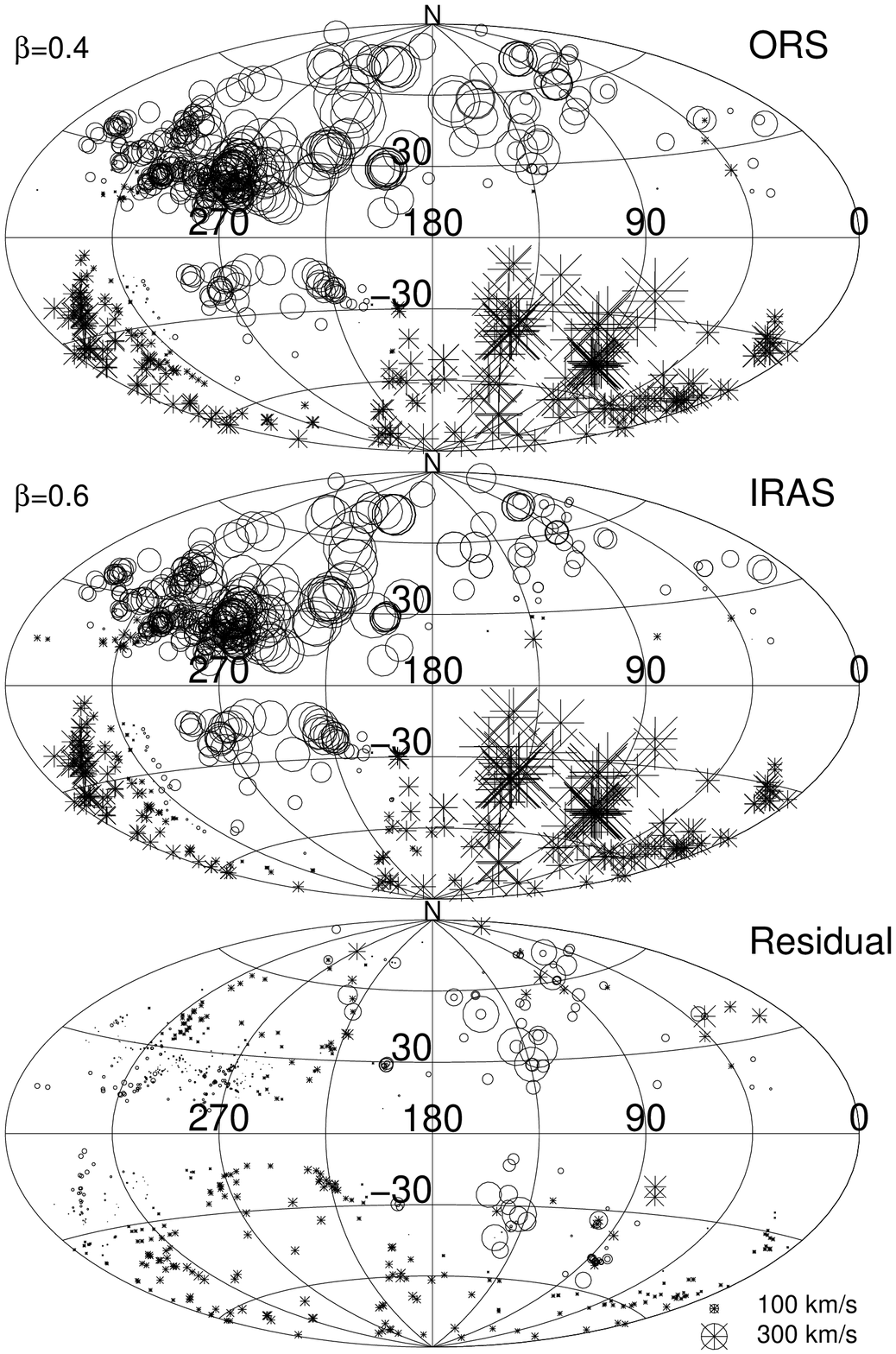}
{As in Figure \ref{vpec1000}, for the middle redshift slice
($2000<cz<4000\kms$).
\label{vpec3000}}

\clearpage

\rsplot{0.9}{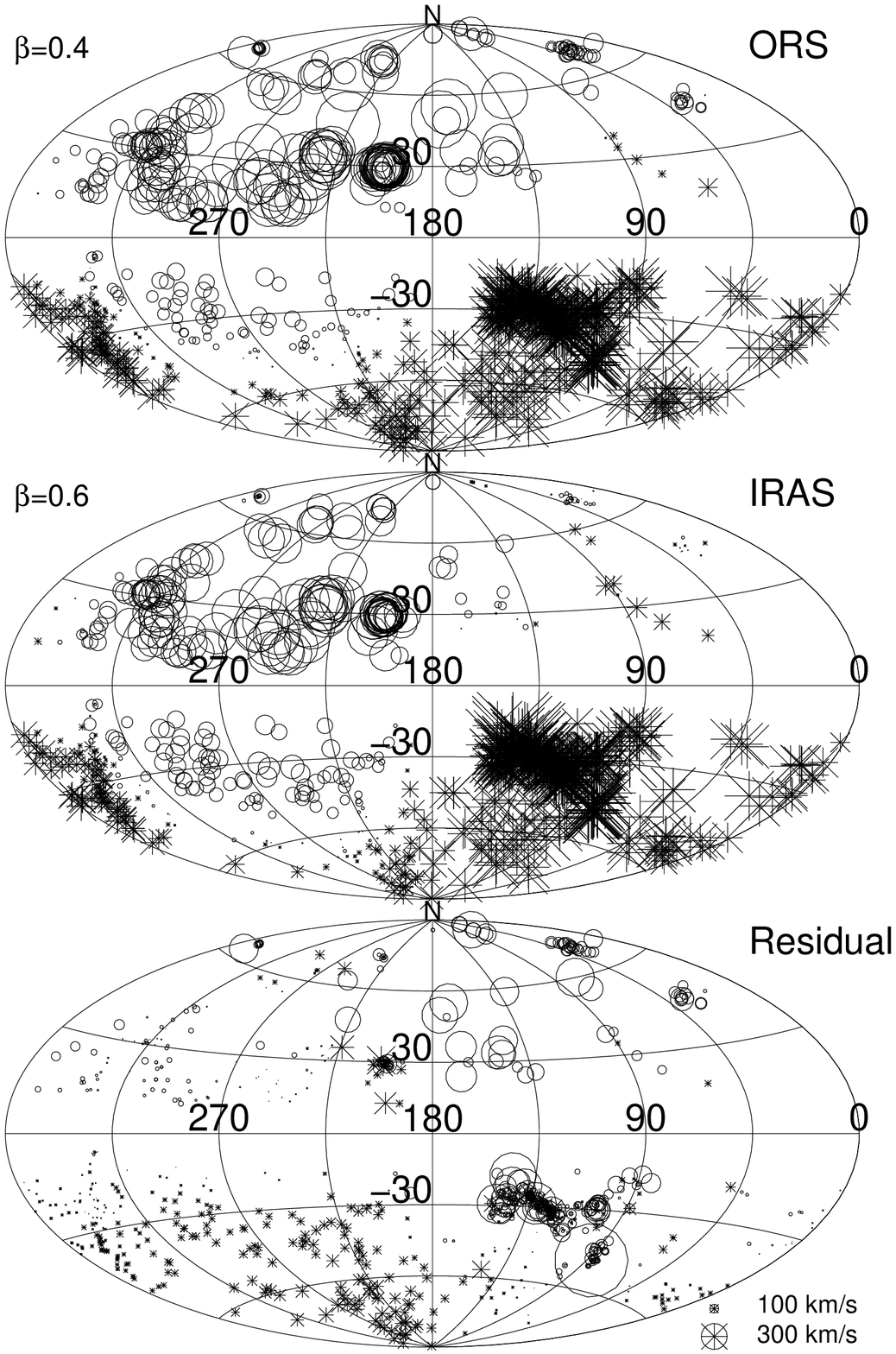}
{As in Figure \ref{vpec1000}, for the distant redshift slice
($4000<cz<6000\kms$). 
\label{vpec5000}}

\clearpage

\rsplot{0.9}{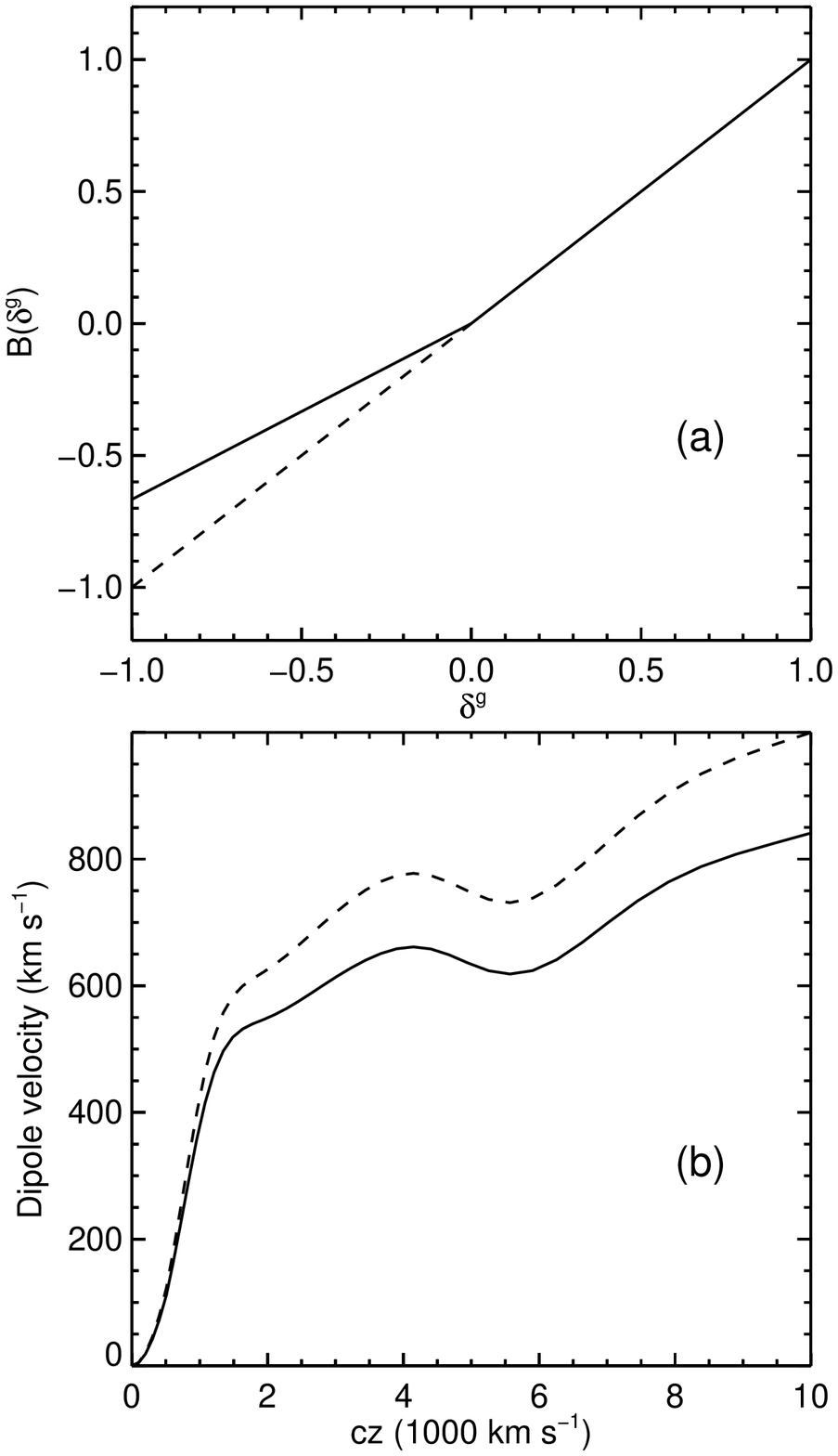}
{(a) A simple non-linear form for the bias function ${\cal B}$ (solid line),
consistent with our variable smoothing procedure.  The fractional fluctuation
in the galaxy density is $\delta^g$.
(b) Effect of the non-linear bias above on the dipole amplitude of the ORS
peculiar flow field with $\beta=0.6$.  The dashed line is for linear bias.
\label{nonlin}}

\end{document}